\documentstyle[12pt]{article}
\begin{document}
\title{Coupled Maps with Growth and Death: An Approach to Cell Differentiation}
\author{
        Kunihiko KANEKO \\
        {\small \sl Department of Pure and Applied Sciences}\\
        {\small \sl University of Tokyo, Komaba, Meguro-ku, Tokyo 153, JAPAN}
\\}
\date{}
\maketitle
%\pagebreak
\begin{abstract}

An extension of coupled maps is given which allows for
the growth of the number of elements, and is inspired by 
the cell differentiation problem.  The growth of elements is
made possible first by  clustering the phases, and then by 
differentiating roles.  The former leads to the
time sharing of resources, while the latter leads to the
separation of roles for the growth. The mechanism of the differentiation of
elements is studied. An extension to a model
with several internal phase variables is given, which
shows differentiation of internal states.
The relevance of interacting dynamics with internal 
states (``intra-inter" dynamics) to biological problems is
discussed with an emphasis on 
heterogeneity by clustering,
macroscopic robustness by partial synchronization and
recursivity with the selection of initial conditions and digitalization.

\end{abstract}

\section{Introduction}

Coupled map lattices (CML) have been used to study many
diverse phenomena of spatially extended systems \cite{CML}: 
(i)spatiotemporal chaos,
(ii) statistical mechanics of an ensemble of chaotic elements, 
(iii)turbulence, (iv)pattern dynamics,
(v)neural dynamics and applications to information processing,
and (vi) biological network problems.

Although the CML approach has been developed rapidly in
the first four fields \cite{CML-rev} and partly in the neural information 
processing field,
it has not been developed so much in
applications to biological networks.

One of the important merits in applying  CML techniques 
to biological networks lies
in the ability to capture the 
interplay between inter-unit and intra-unit dynamics.
Such ``intra-inter dynamics" seems to be essential to a variety of biological
problems.  In cell biology, there are complex metabolic reaction dynamics
in each unit (cell), which are affected by the interaction among cells.
In neural systems, the viewpoint of intra-inter dynamics must be essential 
to the formation of internal images.
An ecological system also consists of interacting
units with internal dynamics.  A CML gives a simple model
for a system composed of interacting units with internal dynamics,
and thus fits with biological problems  better than a cellular automaton,
where there is no internal dynamics in each element.

Indeed, there have been some studies for cellular
biology adopting this dynamical systems approach.

The importance of temporal oscillations
in cellular dynamics was studied in
pioneering work by Goodwin\cite{Goodwin}.  
Recently, the existence of oscillatory 
dynamics for
cell division processes has been discussed both experimentally and theoretically
in cycline and M-phase-promoting factor \cite{Tyson}. 
On the other hand,  different cell types are attributed to
the coexistence of many attractors by Kauffman\cite{Kauff}
where a Boolean network is adopted for each cellular dynamics.
Starting from such internal cellular dynamics,
cell-to-cell interactions are included to study
the Turing-type pattern formation mechanism\cite{Turing}:
Attraction to different states is
found in a coupled system of Boolean-network-type
differential equations \cite{Glass-Kauff},
while a CML corresponding to  Goodwin-type
oscillators is studied by Bignone\cite{Bignone}.

As for the interactions, these studies
basically adopt the pattern-formation mechanism of the Turing 
instability\cite{Turing}.  The internal dynamics is based either on
stable cycles, or on switching type threshold dynamics
allowing only for fixed points.  Use of
chaos or transient unstable dynamics has not been
discussed.  Another missing factor in these studies
is the change in degrees of freedom by cell division and death,
while selection of the system size and boundary condition
through the dynamics itself is
one important characteristic feature in a
biological system.
These are few of several motivations to introduce the
isologous diversification to be discussed in the next section.
With the theory we try to answer the questions of the mechanism of the 
differentiation process
in conjunction with the growth in cell numbers, and how
cellular memory is formed that is transferred stably through cell divisions.

In general, there is an important missing factor in applying the CML approach 
to study a biological problem.  Indeed, this drawback is common
in the dynamical systems (DS) approach to modeling, which may be
termed the ``separation" problem.
A model constructed using the DS approach consists of
time, a set of states, an evolution rule, an initial condition of the states,
and boundary conditions.  It is generally assumed that these four
sets themselves are separated from each other:  Although the states are 
changed according to the
evolution rule, the set of states itself ( e.g., the number of variables)
is fixed independent of the rule.  The states cannot change the 
evolution rule itself.
Initial conditions and boundary conditions are chosen independently of the
state values and of the evolution rule.
In biological problems, such separation between all these
elements of the model may not be valid, or at the very least the
origin of their separation should be discussed.

Let us first discuss the separation between the set of
states and the evolution rule.  In a biological system, the  
evolution rule itself is formed and changes in connection with
the temporal evolution of the states.  A simple example 
is the change of  the number of variables itself with time:
Let us consider the dynamics of a cell society.
When one considers the chemical variability of cells, we need
a set of variables for each cell.  Then,
the number of variables should change with the cell division and
death.  Or, consider another example:  population dynamics.
There  the emergence of new species 
leads to a change in the  degrees of freedom. Such growth of elements
is also seen in economics, where the number of agents can change in
time through reproduction and extinction.

Another aspect of the separation problem lies in the 
segregation of parameters and variables.  In dynamical systems,
the roles of ``parameters" and ``variables" are predetermined and fixed.
In a biological system such separation may not be possible,
or rather, it is important to discuss how some sets of
variables turn into parameters.

The next important problem lies in the choice of initial conditions
or boundary conditions.  For a cell to grow repeatedly, the initial condition
of its internal state should satisfy some condition.  This initial condition, 
however, is determined by its mother cell's state.  Thus, the initial
conditions of a state, and its evolution are not clearly separated.
Through the evolution of the state, initial conditions are selected that
allow for recursive growth.

Recently, the author and Yomo have proposed a novel scenario for cell 
differentiation termed ``{\sl isologous diversification theory}" \cite{KKTY3}.
The cell differentiation and
developmental processes involve internal metabolic reactions,
which are nonlinear, as well as cell division and death, which lead to 
change of the degrees of freedom of the system.  Thus,  
the study of cell differentiation 
is one prototype of our intra-inter dynamics picture for
biological systems.

The present paper is organized as follows.  In \S 2 we explain
the isologous diversification
theory in terms of a coupled metabolic reaction model.
In \S 3 and \S 4, we study a minimal model of the differentiation process,
given by a globally coupled circle map which allows for a change
in the number of elements (cells).  Cells divide or die according to
the history of the rotation of their phase.  We also elucidate the
mechanism by which the phases of oscillation,
as well as their growth rates  are
differentiated.  In \S 5, we extend the model to include
several phase variables and see how differentiation of elements
is embedded into the internal dynamics.
The paper concludes in \S 6 with a discussion of our results.

\section{Isologous Diversification for Cell Differentiation}

The author and 
Yomo have studied a class of intra-inter dynamical models
which are a dynamic model for the cell differentiation process 
\cite{KKTY,KKTY2,KKTY3}.  
This class of models consists of
a metabolic or genetic  network within each cell, and interaction 
between cells through competition for nutrition and 
the diffusion flow of chemicals to media.
In each cell there is a set of chemical variables. When a cell is isolated,
its chemicals are assumed to show oscillatory behaviors of a 
Lotka-Volterra type, that is, the concentration of each chemical
component switches between a low and high
level periodically in time.  As for the inter-dynamics,  
cells are assumed to interact with each 
other through the media.  The interaction here is global, and 
the system belongs to a class of globally coupled dynamical systems.
The cell is assumed to divide and thus gives birth to a new cell
when a  product of the metabolic chemical reactions
exceeds some threshold.
The concentrations of chemicals of two divided cells
are chosen to be
almost identical upon the division.

Starting from a single cell initial condition, we have
found the following scenario for cell growth and differentiation.

\begin{itemize}

\item
{\bf (1)  Synchonous oscillations  of  identical cells:}
Up to some threshold number of cells, all oscillate synchronously,
and their states are identical.

\item
{\bf (2)  Differentiation  of the phases of oscillations  of  internal
states:}
When the number of cells exceeds the threshold, they lose identical and
coherent dynamics.  The phases of oscillations split into several groups
(clusters).  This clustering is a general consequence
of coupled oscillators (maps) \cite{KK-GCM,CC,Okuda,JJ,Nakagawa}, 
when there is strong interaction among them.   

\item
{\bf (3)  Differentiation  of the amplitudes of internal  states:}
At this stage, the states of cells are different even after taking the 
temporal average over periods.  
The pattern of orbits in the chemical phase space 
differs by groups. The dynamics as well as the average behavior of cells is
differentiated.

\item
{\bf (4)  Transfer  of the differentiated state to their  offsprings  by
reproduction:}
The differentiated character of  a cell is transferred to its offsprings.
This ``memory" is made possible through the
transfer of initial conditions for the chemical variables of
the reproduced new cell.

\end{itemize}

As the cells continue to reproduce, the competitive
interaction among them gets stronger
and leads to successive diversification of their behavior.
Generally speaking, identical elements tend to become diversified
through the interplay of nonlinear oscillations, cell to cell interaction, and
reproduction.
The first three stages listed above
are consequences of globally coupled dynamical
systems.
The emergence of the fourth stage, on the other hand,
is attained only through the  reproduction
of  cells, where the initial conditions are selected so that the
offsprings keep the same character as their mother cell.
We believe that this emergence of recursivity or memory is 
an important feature  of  coupled dynamical systems with 
reproduction, and thus
is essential to the information flow and memory in biological systems.

Of course, the idea to attribute a different cell type to a different state
of chemical dynamics is not new.  As mentioned in \S 1, the
existence of multiple fixed point states in Boolean networks
has been proposed to provide different cell types \cite{Kauff},
while different oscillatory states coexist with the inclusion of
cellular interactions\cite{Glass-Kauff,Bignone}. 
In comparison with these previous dynamical-systems models 
for cell differentiation, however,
the proposal of our theory is novel as to the following points:

{\bf (i) Importance of instability}:
In the second stage, clustering by coupled
nonlinear oscillators is essential to the trigger of differentiation.
For the clustering, the orbital instability of a system is required, either
by internal dynamics or through interactions.  
The dynamics of the attractor itself is not necessarily chaotic, but 
an instability at least during transient time steps is required.

{\bf (ii) Non-diffusive interaction}:
In contrast with the previous theoretical models for differentiation,
the interaction form is not diffusive as postulated for the Turing 
instability, but is based on the (global) competition for chemical resources.
With this interaction, differentiation proceeds.  In particular, this form 
seems to be
essential to the fixation of differentiation (with amplitude clustering) at
the third stage.  

{\bf (iii) Stability at an ensemble level}:
Our differentiation is based on the interaction among
nonlinear elements.  In such a coupled system,
stability of a collective dynamics 
has been studied, that is formed by
an ensemble of chaotic elements \cite{KK-GCM,GCM2,homeo}.  Such stability is
essential to our scenario, and is indeed found in our simulations \cite{KKTY}.
For example, when all cells of some type are removed, other
cells of a different type are transformed into the removed type,
through divisions.  With these changes
of cell types, the cellular distribution
comes back to the original one.
This stability is especially important for the maintenance of a biological 
system.

{\bf (iv) Change of degrees of freedom in conjunction with dynamics }:
Cell division and death lead to the change in degrees of freedom, which
provides a novel class of dynamical systems.
There a novel type of instability is proposed as ``open chaos" \cite{Alife} 
where the
orbital instability in global phase space is in conjunction with the change of degrees of freedom.
Another important factor in such a system is the selection of
system size.  With cell division and death, the variation of the
total number of cells remains within some range.
Such autonomous selection of the system size is important in a biological 
system.
For example, apoptosis is essential to the selection of the total 
number of cells in a system,
and our model may give a conceptual model of it.

{\bf (v) Recursive transmission through selection of initial conditions }:
The cellular memory at the fourth stage is formed as the result of the selection of
initial conditions for a cellular state (i.e., a partial system of the total 
dynamical system).  This argument is possible only for a system with a
cell division process, internal dynamics, and interactions.  As for the choice 
of initial conditions of the internal cellular system, this selection could be  related with the
basin for multiple attractors. However, in our model, the cellular interactions
are also relevant to the formation of multiple cellular states and the 
selection of one of them.
The coexistence of multiple states and the selection are also the outcome of
both the interactions and internal dynamics.

\section{Coupled Map Model with Division and Death: Relation with Growth and Synchronization}

The cell differentiation model adopted in \cite{KKTY2,KKTY3} is
rather complicated in order to correspond with biochemical reactions.
In this section we consider a very simple, possibly the simplest, model
with internal oscillatory dynamics, competitive interaction,
and growth in the degrees of freedom.

First we assume that there is a variable, $x(i)$, determining the cell state,
and that cells compete with each other for a source term $s$.
The source is supplied from the outer environment with a constant rate $s$.
The ability to get this source depends on the internal state $x(i)$.
Thus the dynamics of each $x(i)$ is given by

\begin{equation}
x_{n+1}(i)= x_n (i)+ f(x_n (i)) + S_n;
\end{equation}
\begin{equation}
S_{n} = \frac{s - \sum_j f(x_n (j))}{N}.
\end{equation}

The term $x_{n+1}(i)- x_n (i)= f(x_n (i)) + S_n$ gives a source term
that the element $i$ takes at the time step $n$.  The second condition
assures $\sum_i \{ x_{n+1}(i)- x_n (i) \} =s$, that is,
the sum of the source term balances with that supplied externally.

Since $x_n(i)$ represents an internal state, which is oscillatory,
it is natural to relate it with the phase of oscillation.  
This  correspondence inspires us to choose the periodic
function of  $f(x)=\frac{K}{2\pi} sin(2\pi x)$,
so that the dynamics depends only on the phase, represented by the
fractional part of $x(i)$.
It is useful then, to relate the threshold for division 
with the number of cycles after the previous division.
Hence we assume that the condition for division and 
death is governed by the the number of oscillations, given
by the integer part of $x_n(i)$.  Taking these considerations into account,
we choose the following rules for the division and death:
 
(A) Divide cell $i$ if $x_n(i)> T_g$; After division, $x_n(i)$ of 
cell $i$ and the new element $N+1$ is assigned to be
 $(x_n(i)- T_g)/2 +\delta$  and $(x_n(i)- T_g)/2 -\delta$, respectively,
with $\delta$ a very small random number.

(B) Remove cell $i$ if $x_n(i)< T_d$;
 
Here $T_g \geq 1$ and $ T_d \leq 0$ are integers giving the threshold for
growth (division) and death, respectively.
We often call element $i$ as cell $i$, and the
duration from its birth (or division) to the next division ( or death) as
the cell's lifecycle, following the analogy with  cell biology.
The number of time steps from its latest division (or birth) to the
next division ( or death) is called its lifetime.
Although no difference exists besides the error term between the two 
cells $i$ and $N+1$,
produced by the division of the cell $i$, one of them 
remains to be called cell $i$.
Since each cell $i$ can divide several times, we attribute a lifecycle and 
lifetime
to each division process, here.  (Thus there can be several lifetimes
for cell $i$ according to successive divisions).

The model of (1) is a globally coupled map (GCM)
with the coupling term given by (2).
Thus it is expected that the dynamics of (1) leads to the
clusters of synchronized oscillations
as in globally coupled circle maps\cite{CC}, which
are relevant to the dynamics of growth.  Here a cluster is defined as a set
of elements having identical fractional parts of $x(i)$ up to
a prescribed precision.  The integer part of $x_n(i)$ is not taken into
account for the definition of clustering, since the dynamics (1) only depends on 
the fractional part, and two elements with the same fractional part show 
identical
oscillations until the division occurs to which the integer part
is relevant.  In other words, the dynamics of the phase variable $x(i)$ 
is identical up to this precision for elements belonging to
an identical cluster.  If all elements belong to a single cluster, 
they are synchronized perfectly.
Since the integer part of $x_n(i)$ can be different 
even for two cells belonging 
to the same cluster,
the division condition can be applied to them at different time steps.
(Note that the history of two elements (such as the time of the latest 
division)
can be different even if two elements are synchronized at the moment).
Through this division process, two completely synchronized elements
(even up to an infinite precision) can change their values.
Thus the clustering condition itself is dynamic, in contrast with
the clustering of an attractor in GCM \cite{KK-GCM}.

In the present paper we define a cluster at each time step
with a precision of $10^{-5}$.
A state of an ensemble of cells is
classified by the number of synchronized
clusters $k$ and the number of elements for each cluster $N_k $,
i.e., the partition of $N$ elements to $k$ clusters
as $(N_1 ,N_2 ,\cdots,N_k )$.

Here  we mainly discuss the simulations with $T_d=0$.  In this case,
the number of cells does not grow indefinitely.
As $K$ is increased, we have seen roughly three phases:
\footnotemark

\footnotetext{ see \cite{Rel} for the discovery of these phases
for $T_g=1$.}

(1) Ordered Phase

(2) Partially Ordered Phase

(3) Desynchronized Phase

Each phase is characterized as follows
( see Fig.1 for the time series of $x_n(i)$ for each phase):

(1) Elements tend to be synchronized.  As shown in the time series
(see Fig.1a),
elements' oscillations split into two clusters for some interval,
and then all of them tend to be synchronized, and 
later split into two ( or a few)
clusters.  This process repeats with time as $x_n(i)$ increases, 
within each cell's lifecycle.  As the number of cells increases,
their oscillations increase the mutual coherence, in the present model.
When the number gets larger, all cells' oscillations become coherent.
Then there appears simultaneous deaths of multiple cells,
and the number of cells decreases drastically, from which
the growth again starts. (see Fig.2a for the change of the total number of 
cells $N$).
Thus the system has two levels of cycles; one is the intracellular oscillation,
and the other the intercellular oscillation associated
with the change of the number of cells.

(2) Partially ordered phase :
Around $K \approx 3.3$ the oscillations of identical cells can be 
desynchronized.
The cluster number fluctuates between 1 to $N$, while the coupled system 
consists of a single large synchronized cluster ( i.e., $N_1 \sim N/2$)
and many other desynchronized elements.
Here the number of cells can grow up to a very large number.  
Indeed, as is seen in Fig.2c,
there are two temporal regimes; for most of the time the number oscillates
around O(10), but occasionally there appears an intermittent burst to a 
very large number
of cells ($100 \sim 700$).  The timeseries in the former regime is given
in Fig.1b, while that for the latter case is given in Fig.1c.
In this partially ordered phase, the number of cells 
often stays at a large value for
about a million steps, until there is a simultaneous death of 
many cells.  After the death of many cells,
new cells start growing as in Fig.1d.
Detailed study of this phase is given in the next section.

(3) Desynchronized states:  For $K>3.4$, elements' oscillations are
typically desynchronized.  For most time steps,
all elements are desynchronized with each other, i.e,
the cluster number is $N$ (see Fig.1e, where the number of cells
fluctuates between two and six) .
Growth of the number of cells is suppressed.  The number fluctuates
at a low level, with an irregular oscillation.

To see the above changes quantitatively,
the maximal and average numbers of cells are plotted as
a function of $K$ in Fig.3.  
There is a sharp peak at $K \approx (3.2 \sim 3.3)$,
independent of the choice of threshold $T_g$.
At $3<K<3.4$, the maximal
cell number is over a thousand ( see Fig.1), which is due to the existence of
a temporal regime allowing for a steady increase to a large number
and its maintenance (see Fig.2c).  In Fig.4
we have also plotted the average fraction of cluster numbers, i.e,
$<k>/<N>$, where $< >$ is the temporal average.
There is a sharp increase 
at $K\approx 3.3$, beyond which the fraction is close to 1,
meaning complete desynchronization.

To see the instability of synchronization, it is useful to introduce
the split exponent following the definition given in
GCM \cite{Inf}.  It is defined as
the rate of amplification of $x_n(i)-x_n(j)$ of two elements
such that $x_n(i)\approx x_n(j)$.  Since the global interaction term
is common to all elements, the exponent is given by 
the average of the expansion rate for the  
one-body part of eq.(1), i.e., $x_{n+1}=x_n(i)+f(x_n(i))$.
Thus it is defined as

\begin{equation}
\lambda_{spl}(i;T_0,T)= (1/T)\sum_{n=T_0}^{T_0+T}log |1+f'(x_n(i))|.
\end{equation}

The above exponent is an average over time steps $T_0$ to $T_0+T$.
If the element remains existing forever, it is possible to 
take the infinite time limit  to obtain a well-defined quantity
like the Lyapunov exponent, as long as we do not take into account
the perturbation caused by the ``division process", which is 
not represented by the mapping process.
In our problem cells can divide or die, where a threshold-type instability
sets in.  Still, the following quantifiers should be relevant to discuss
the split instability:  
(a) The average of the exponent over all cells and over all time steps --
this quantity, denoted as $\lambda_{spl}$, measures the average tendency of 
desynchronization.
(b) The average of $\lambda_{spl}(i;T_0,T)$ over a cell's lifecycle
(i.e., from its latest division (or birth) to its next division (or death)) --
this quantity is obtained just by taking $T_0$ as the 
latest first division time
(or birth) and $T+T_0$ as the next division time ( or death).
The quantity measures an average degree of synchronization of a cell
over its lifecycle.

The change of the average split exponent $\lambda_{spl}$ versus
$K$ is given in Fig.5.   
The exponent becomes positive around $K\approx 2$.
We note that the growth of cell numbers
starts to increase at $K>2$.  Between $2<K<2.9$, which corresponds to the
ordered phase,
the average exponent remains close to zero, as the 
``synchronization to few clusters" and ``split by division"
are balanced.  This balance is shown in Fig.1a, where the above two processes
repeat in each cell's oscillation  within each lifecycle.
For $K>2.9$ the exponent starts to be positive,
and increases slowly thereafter with $K$, until
$K\approx 3.3 $, where there is a sudden jump in value.
The regime $2.9<K<3.4$ corresponds to the partially ordered phase.
For $K>3.4$, i.e., at the desynchronized phase, the exponent smoothly 
increases with $K$.

The internal dynamics differs between cells that succeed in division,
and those that will die.  In Fig.6, we have plotted the histogram of
the split exponent over a cell's lifecycle for the cases of division and death,
while the split exponent versus the cell's lifetime is given in Fig.7.
As in Fig.7b, most cell deaths occur within a few steps after a cell's
latest division.  Such ``quickly" dying cells produce a peak in the histogram
for $\lambda_{spl}>1$ in Fig.6b, while those cells that died $10$ steps
after their latest division produce a peak around 0.3.  
Some dividing cells also produce a broad peak around 0.3,
while those cells dividing after 100000 steps
lead to a different peak of the exponent, around .15,
as shown in Fig. 6a and 7a, respectively. In Fig. 7a, we note that there
are two groups of cells with differing split exponents.
This grouping suggests that cells with long lifetimes have become 
differentiated into a type with very long lifetimes
and that with less long ones.
Indeed such long-living cells appear while the
number of cells is large as in Fig.2c).
We will discuss the mechanism of this differentiation in the next section.

As is seen in Fig.3a, the growth of cells is enhanced in the partially
ordered phase where elements are partially synchronized.
It should be noted that the maximum rate of growth occurs not at 
the marginal stability point ($\lambda \approx 0$), but in the regime with
a small positive  value, in contrast with the ``edge of chaos"-type
picture.

If the oscillations of all the elements are synchronized,
then all elements compete for the source term at the same time.
This intense competition does not allow for an effective use of resources.
Instead, through the clustering of elements into different groups, 
a sort of time sharing system is constructed.
Thus resources are effectively used, because there is some ordering of the
elements.
In the partially ordered phase, the ``roles" of the elements are also
differentiated.  As will be discussed in the next section, 
cells in large coherent clusters stop dividing but remain existing without
dying,
while other elements' oscillations are unstable and proceeding to grow faster.
For larger $K$, elements are  completely  desynchronized, and
no ordering for the use of resources is possible.
In this case effective use of resources is once again impossible.

Before discussing the detailed dynamics of the system,
let us briefly mention the dependence of our dynamical behavior
on $T_g$ and $T_d$.
First, the essentially same behavior is observed for $T_d=0$, independent of
$T_g$.  The three phases are found with an increase of $K$. The
dynamics of $x_n(i)$ and the number of cells are
basically the same as we have discussed above.

Second, when $T_d$ is negative, cells can grow indefinitely in number,
according to our simulation results so far.  This is because  in 
our system, $s$ is
constantly supplied, and cell's number can increase with the rate
$s/(NT_g)$ on the average as long as the death condition is
not satisfied.
When $T_d$ is negative, the death condition is not
satisfied if $K$ is small, where the growth seems to
continue indefinitely.
The growth rate is slightly enhanced  as the desynchronization is increased,
up to $K\approx 3.5$, where cell death starts to set in
and limit the growth.\footnotemark

\footnotetext{  At $T_g=1$, however,
there is an explosive growth for
 $K\approx 3.5$, where some deaths of cells effectively enhance others'
growth, since removal of negative $x(i)$ by death
adds up the source term.  This is an artifact in our model,
since the conservation of the source term is not
satisfied when a cell dies.}

\section{Growth with Differentiation of Roles}

Here we study the dynamic origin of the sudden change from a steady
regime with a small number of cells to the explosive growth in the number,
seen at the partially ordered phase as in Fig.2c.  As an example,
let us consider the case given in Fig.8.
Around time steps $4 \times 10^5$ to $6 \times 10^5$,
the number of cells increases to a large value.
This regime is clearly distinct from other regimes.  
As seen previously in Fig.2c, such growth regimes appear
intermittently.

In Fig.9,the split exponents of divided (Fig.9a) and dead (Fig.9b)
cells are shown
versus time, while the cluster sizes of divided and
dead cells are plotted in Fig.10.  Here the cluster size of a cell
is defined as the number of cells which have an identical fractional
part of $x_n(i)$ up to the given resolution.  
One can see clearly that the nature of divided and dead cells in this
temporal regime, as characterized by $\lambda_{spl}$, is different 
from that in the other regime.
From these results and
the direct measurement of the time series of $x_n(i)$, 
the growth regime is characterized as follows:

(i) cells split into two groups.  One group of cells forms
a large cluster (typically on the order of $N/2$), while the other
cells are mostly desynchronized with each other, although 
they can form a synchronized cluster of a smaller size
and collapse intermittently.

(ii) The former group of cells neither divides nor dies.
In this temporal regime,
only groups of cells that are desynchronized 
with each other divide and lead to growth in the total number of cells.
This desynchronization of  divided or dead cells
is clearly seen in the time series of their cluster sizes.
For example, as Fig.9 shows, there is a remarkable lack
of clusters larger than size 10
around the time steps $(5-6)\times 10^5$.
In Fig.10, cells with large and small split exponents
are lost during that regime, which corresponds to the fact that
the elements in the synchronized cluster are no longer dividing.

(iii) The collapse of the growth occurs 
because of the synchronous division of many cells belonging to the large
cluster.  It is clearly seen in Fig.10 that cells belonging to
the cluster divide and die successively at around the $6.2\times 10^5$ 
time step.
We have also measured the time between
divisions for each cell (i.e., cell lifetime) .  This time fluctuates around 
$(1-10) \times 10^3$ 
for most time steps, but around time step $6.2\times 10^5$,
it suddenly jumps to $(20-300) \times 10^3$,
meaning that the cells that did not divide during the growth regime,
have successively divided.
With this division and the subsequent simultaneous deaths of many
cells, the number
of cells is drastically reduced back to the normal level.

It is rather interesting to note that
the formation of  two distinct regimes 
(i.e., sudden growth and normal growth) is 
supported by the emergence of the two distinct groups of cells.
The group of cells forming the large cluster is necessary to
support the growth of the other group which consists of desynchronized cells.
The growth of the latter group, on the other hand, suppresses
the growth of the former 
and maintains the stability of this cell society with its inhomogeneous
clustering.  The differentiation of the roles that the two
groups play in producing growth of new cells and maintaining the stability 
of the system makes 
possible the growth regime.  

Following this separation of roles, it is interesting to
draw a cell lineage diagram.  
In Fig. 11, we have plotted this diagram, where
the division process
with time is represented by a horizontal line between
mother and daughter cells, while a line is terminated when the
corresponding cell dies.   The diagram shows the differentiation of
cells as to the number of offspring, as well as the successive appearance of
multiple simultaneous deaths.

\section{Differentiation of Internal States in Coupled Circle Maps}

\subsection{Multi-phase Model}

In the model in the previous sections, the internal dynamics was 
represented by only a single
phase dynamics.  Since the single circle map has one variable and one attractor
for most parameter regimes, the 
different dynamic behaviors appear only through clustering relationships with 
other cells.
Indeed the different behaviors of a cell are governed by 
the size of the cluster it belongs to, rather than 
by its internal states.
Thus it is necessary to study a model with internal variables rather than a
single phase variable, in order 
to study the fixed differentiation memorized in the
internal states of a cell.

The author and Yomo have studied a model with
biochemical reaction dynamics \cite{KKTY,KKTY2,KKTY3}, 
given by a set of ordinary differential 
equations with switching-like oscillatory dynamics.
There the amplitude of the oscillation of chemical concentrations in a cell
is essential to the fixed differentiation coded in an internal state
of a cell.

Here we consider briefly a simpler model with several phase variables within
each cell.
Assume that the internal dynamics is composed of several
cyclic processes, which are each represented by a circle map.
The internal dynamics, given through the interaction among the cyclic 
processes, is just represented by a coupled circle map:

\begin{equation}
x_{n+1}^{m}(i)= x_n ^{m}(i)+ f(x_n^m (i))+
\sum_{\ell}a^{\ell,m}g(x_n^m (i)-x_n ^{\ell}(i))+S_n^m;
\end{equation}

\begin{equation}
S_n^m = \frac{s^m - \sum_j f(x_n^m (j))}{N},
\end{equation}

for $m=0,1,2,\cdots, M$.  This superfix $m$ corresponds, for example, to
each metabolic chemical cycle involved in each cell, or generally
speaking, to some internal process.  Here we call
this superfix $m$ a {\sl chemical species} for simplicity.

We choose the source term $s^0>0$, and $s^m=0$ for $m>0$, 
assuming that the component $0$ is
the source chemical (e.g., nutrition), while the other components' cycles
correspond to the metabolic processes that bring about the growth of the cell. 
Taking these allocations into account, we define the 
condition to grow and divide by
$\sum_{m=1}^M Int(x_n^{m} (i))>T_g$ while
the death condition is defined by  $ Int(x_n^{0} (i))<T_d$, where
$Int(z)$ denotes the integer part of $z$.
When a cell divides, a new cell is created with
chemical species concentration
$x_n^m (i)-Int(x_n^m (i))- \delta ^m$ where $\delta ^m$ is a small random number
(taken from a uniform distribution over $[ -10^{-5},10^{-5} ]$),
while the original cell has chemical concentrations $x_n^m (i)-Int(x_n^m (i))+\delta ^m$.\footnotemark

\footnotetext{Here we have not divided the fractional part of $x_n^m(i)$ into
two, in contrast with the model in \S 3.  Indeed the choice is
rather arbitrary.  Qualitatively identical results are obtained even if
half of $x_n^m(i)-Int(x_n^m(i))$ is transmitted at the division.}

The coupling terms $a^{\ell,m}$ and the function $g$ represent the
interaction between the cyclic processes.  Taking into account
the periodic nature of these processes, we adopt
again sine circle maps $f(x)=\frac{K}{2\pi}sin(2\pi x)$ and  
$g(x)=\frac{c}{2\pi}sin(2\pi x)$ to model them.
Since the coupling term represents the flow of chemical process,
it is postulated that $a^{\ell,m}=-a^{m,\ell}$.
First we assume that $a^{0,m}=c$ to assure the flow from
the source chemical to other species.
Next, for other coupling coefficients $a^{\ell,m}$
we set most of them to zero, but
leave a few of them at a constant $c$.   Such pairs $(\ell,m)$ that give 
$a^{\ell,m} \neq 0 $ are randomly chosen with the rate $L$ 
per chemical species.
In the present section we take $M=8$, and $L=2$.
In other words, we have chosen a model with sparse connections.
We have made several simulations for this particular coupling
sequences of the $a^{\ell,m}$ with this particular $L,M$ condition.

\subsection{Differentiation in Chemical Compositions}

For most parameter regimes, no growth in the number of cells is observed.
We have found that either all cells die out or
cells stop increasing at a small number ( e.g., from 2 to 8).
This is because no flow from
the source $x^0$ to the other variables is formed,
and thus the growth condition $\sum_{m=1}^M Int(x_n^{m} (i))>T_g$
is not satisfied.  When division stops, chemical 
oscillations of an individual cell
are synchronized across all cells.  Hence, time sharing for 
resources is not attained, and further growth is suppressed.

Only at $ 0.59 \leq c \leq 0.62$ and $ .8 \leq K \leq 1.1 $, have we
found successive growth and death processes.
Here we survey the dynamics in this region.

First we note that all chemical oscillations are desynchronized from cell to cell
as the number of cells increases.
This desynchronization appears at a rather early stage ( after one or two
divisions).  The  temporal  average of chemicals starts to differ later,
after a few divisions.
Here again, clustering is essential for
steady growth; otherwise no growth is observed as in most other parameter
regimes.  

An example of the time series of the number of cells is
given in Fig.12.  In the figure, there are
clearly two distinct types of temporal regions;
those with and those without a frequent change in the 
numbers of cells.  In Fig.12, the latter regime is seen e.g., 
around $3 \times 10^{7}$ 
and $5 \times 10^{7}$ steps.
Such multiple temporal regions are typically observed  at the parameter region
near the boundary for growth (such as $c \approx .59$ or
$K\approx 1.1$), while regimes without the growth do not contain clearly 
separated regions (for example at $c=.6$ and $K=1$).

To see the average property of a cell, we
define 

\begin{equation}
R^m_n(i) = Int(x^m_n(i))/(n-T_b),
\end{equation} 

where $T_b$ is the time of the latest division ( or birth)
of the cell.  In other words, $R^m_n(i)$ measures the average
rotation of the chemical cycle $m$ at the cell $i$, per step.  The rotation
$R^m_{T_d}(i)$ at the next division time step $T_d$ gives the
rotation rate over the lifecycle of the cell.
This rotation $R^m_n(i)$ gives the contribution of each chemical species to
the process of division.  This quantity gives a measure
for the activity of each chemical cycle, or
roughly speaking, each chemical composition.

In Fig.13 we have plotted $R^m_n(i)$ 
for several time steps $n$ ((a)-(d)),
against the cell index, while the rotation $R^m_{T_d}(i)$ at the division
is plotted against time in Fig.14.
On the average, the $M$  chemical species split into two groups,
one for $R^m_n(i)>0$ (which has two species in the figure), 
and the other for $R^m_n(i)<0$
or slightly positive (which has the other 6).  
In other words, the chemical species become differentiated within each cell.  

The above ``chemical"
differentiation applies for each cell.  Besides this intra-differentiation,
cells are separated into several  groups as well, as shown in Fig. 13.
In Fig.13a, the rotation $R^m(i)$ for each $m$ does not differ by cells so 
much.
As for the average chemical compositions, all cells are almost identical,
although the phase of oscillations itself is not synchronized.
With time, cells with differently behaving $R^m(i)$ appear as in Fig.13b.
Roughly speaking, there are two types.  In one type,
the difference of rotations between two chemical groups
( with 2 and 6 components respectively) is much larger; for this type of cell,
positive values of $R^m_n(i)$ for two chemicals are much larger than the other 
type and the negative values for the other 6 chemicals are smaller.
In the other type of cell,
the difference between the rotations of two groups of chemicals is much smaller.
Often the sign of $R^m_n(i)$ is opposite, that is,
6 chemical species have slightly positive $R^m_n(i)$, and the other two have slightly negative ones.

We note that to satisfy the division condition, the sum of $R^m_n(i)$ over
chemicals $m$ must exceed the threshold.
Thus the above two types correspond to two strategies to satisfy the threshold condition:
one is to have few chemical species of large positive rotation values, while
the other is to keep the magnitude of negative rotation values smaller.
The former cell is chemically specialized, while the latter
cell sustains chemical diversity, in the sense that all chemical
cycles contribute to the growth.
Differentiation of cell types is associated with that 
of chemicals contributing to the growth.
With this differentiation, the cells' competition for the chemicals
is reduced, which is relevant to the growth.

This differentiation is not a snapshot property (i.e., depending on
the phase of oscillation), but the average property over a cell's lifecycle. 

With a cell's division,
the average rotation  $R^m_n(i)>0$ may switch its sign.
Indeed the rotation $R^m_n(i)>0$ oscillates with time $n$ through the division.
Hence the types of a cell may change with division.
To see if the differentiated type of a cell is recursively transmitted,
we have plotted the return map of $R^m_{T_d}(i)$ versus $R^m_{T_d}(j)$
for the cell $i$ born from the cell $j$ by division.
If complete recursivity held, the plot would
lie on a diagonal line.  Starting from a randomly chosen 
initial condition, there
is an approach to recursive transition, 
but it is attained very poorly.  The memory is not preserved as in the
model in \S 3.

The rotation $R^m(i)$ stays only within some finite range.
In Fig.14, a majority of cells are
recursive around $(-2 \sim -1)\times 10^{-3}$,
while cells deviated from this region are less recursive.
Some of the deviated cells around $(.5 \sim 1)\times 10^{-3}$
remain there, but most of them lose this characteristic at their
next division.
As in Fig.14, there is no point along the diagonal region for highly 
deviated cells, which means that such deviation is only 
a one-generation property not transmitted to daughter cells.

\subsection{Temporal Switch of Stages} 

In Fig.12 and 13, we
have seen that there are two distinct temporal regimes as to the growth patterns.
In Fig.15, we have plotted
the rotation $R^1_{T_d}(i)<0$ (over a cell's lifecycle)
versus its time of division, 
to see the difference between the regimes.
There, a larger fraction of cells has $R^1(i)_{T_d}<0$, while cells with the
opposite rotation appear with a smaller rate.
It should be noted that Fig.12 and 13 suggest that
the birth of cells is more frequent
when the value $R^m_{n}(i)$ is scattered by cell $i$.
If the distribution of $R^m_{T_d}(i)$ over cell $i$ is
concentrated, division is suppressed.  For example,
around time step $3\times 10^7$,
a single type of cell dominates as in Fig. 13c.  That is, 
$R^m_{n}(i)$ for all $m$, takes almost all the same values for cell $i$.
Around  this time step,
growth is inhibited, as is seen in Fig.12 (see also Fig.15).
As in Fig.13d and Fig.15, growth is sustained by the heterogeneity of the
cell society.

We have also measured the histogram of $R^m_n(i)$ 
over all cells for some time intervals. 
The distribution has two broad-band peaks 
during the time course when cells are dividing  frequently,
although more than $3/4$ of the cells are
accumulated in one peak (at $R^1(i)<0$).
Cell division stops when the distribution has one (broad) peak.
The coexistence of different types of cells
seems to be necessary for the overall growth of cells.

Summarizing several simulations, the cell society
evolves as follows:  As cell types get similar
(e.g., with the same sign of $R^m_n(i)$ for all species $m$),
cell division is suppressed.  
A homogeneous cell society, however, is unstable,
and the cells' small differences start to be amplified.
A few cells start to be 
differentiated (i.e., to have a different sign of $R^m_n(i)$)
when growth is enhanced.  Then cells divide several times,
followed by some cell deaths.
Successive changes between
a rather homogeneous cell society, and a highly heterogeneous society
allowing for  growth are observed repeatedly.
It should be noted that the differentiation of different cell types is
not just by their phase of oscillation,
but by the change of chemical species contributing to
the condition $R~m_n(i)$.

\section{Discussion}

In the present paper, we have studied a simple coupled map model
inspired by cell biology.  In a biological system, there always exists
an intrinsic tendency toward diversity and individuality.  A biological 
system is essentially heterogeneous, as is simply
seen in the fact that no two cells are identical, in contrast with 
physical particles like electrons which are all exactly the same  \cite{Elsasser}.
In terms of dynamical systems,
this diversification is purported to be supported by the
orbital instability of the evolution of a cellular state
provided by a chaotic system \cite{Rel,Alife}.
As is seen in the clustering phenomena in coupled map systems,
elements split into groups with different dynamical behaviors.
Assuming such instability exists at the individual cell level, 
an important question is how macroscopic robustness is sustained.

Our hypothesis here is that the stability is sustained at an ensemble level.
Coupled nonlinear systems have a rich
variety of collective dynamics, ranging from low- to high-dimensional
behavior \cite{Nakagawa,GCM2,Pikov,Perez1,CM}.
Furthermore this collective dynamics often maintains stability even if it
is high-dimensional chaos.

In the present simulations, it should be noted that a system keeps 
partial synchronization to sustain 
the growth in the number of elements.  When growth is possible, the system
bounces between synchronization and complete desynchronization.
When synchronization dominates, multiple simultaneous deaths of cells 
occur, and the system restores diversification having different
oscillatory dynamics.  When desynchronization dominates,
on the other hand, no cooperative use of resources is possible. There is
too much competition for resources, and cells die before
increasing their number.
The relevance of partial synchronization is clearly seen in the growth regime,
where the differentiation into a small group of 
rapidly dividing cells and a majority group of
non-diving cells  makes possible the overall growth of cells.
In the regime where cells are able to grow, the split exponent is 
slightly positive.

In a model with an internal coupled circle map, a few types of cells
are formed with different internal phase rotations.
This variety, to some degree, is necessary for growth; when it is lost 
the speed of cell divisions is lowered, until diversity is 
recovered after the increase of cell number due to division which
changes the nature of the chemical dynamics in each cell.

Summing up, the macroscopic stability of the growing state is sustained by
partial synchronization supplied by a change in the number 
of degrees of freedom (i.e., number of cells).
With this partial synchronization,
the chaotic instability of the whole chemical dynamics
is kept rather weak although many degrees of freedom
are involved.  In an unrelated ecological model, the
author and Ikegami have proposed the notion of ``homeochaos" \cite{homeo} 
which is stability at the macroscopic level supported by high-dimensional
weak chaos, where chaotic itinerancy\cite{KK-GCM,Tsuda-CI,Ikeda}
over several clustered states
is found.  It should be noted that the triplet structure there --
``weak chaos", ``partial synchronization", ``many degrees of freedom" --
necessary for stability is common to our model.
The main difference here is the mechanism that sustains
the partial synchronization.  In our model here, it is due to the change in the
degrees of freedom, while in the model of \cite{homeo},
it is sustained through the change of mutation rates.\footnotemark

\footnotetext{ Although the number of variables itself does
not change in the model of \cite{homeo}, the effective degrees
of freedom changes with the mutation rate, since the populations of many 
``species", whose number changes with the mutation rate,
vanish in the model.  Thus the correspondence with the 
present model
may be even stronger than the statements above.}

There remains an important question from our standpoint, that is, how
the recursive transmission of cell type is attained in a dynamical system 
approach.
Since we have demonstrated the tendency of cells to become more diversified,
the origin of
faithful self-replication is a nontrivial question.
Our hypothesis here is that the ``digitalization" of states through clustering,
and the choice of initial conditions leads to recursive transmission.
A cell with a given differentiated behavior transmits its
character through its initial condition.  In the simulations in \S 5,
a cell's character is not well transmitted to its daughter cell, but
some extension of the intra-inter dynamics may lead to 
the recursive transmission of a mother cell's characteristics to its 
daughters.
Indeed, in the recent model of the author and Yomo \cite{KKTY3}, such 
recursivity is partially accomplished.  In that model, each internal dynamics
has degrees of freedom not only for the phase but also for the amplitude.
The internal dynamics, by itself,  can show an on-off type
switching oscillation, which
may be relevant to producing recursivity,
since a state is easily coded digitally with the switching-type oscillation.

In short, our dynamical systems viewpoint of biology is summarized as follows:

{\bf

1) heterogeneity by isologous diversification $\Leftrightarrow$ clustering 
due to orbital instability in each internal dynamics

2) macroscopic robustness $\Leftrightarrow$ collective dynamics of coupled
nonlinear elements, in particular homeochaos

3) recursivity  $\Leftrightarrow$ choice of initial conditions with 
digitalization of differentiated states  

}

\vspace{.2in}

Although our model is much too simplified, 
it provides a starting point for the study of
biology from the above standpoint of dynamical systems theory.
Besides the dynamic origin of a cell differentiation,
our results suggest the relevance of heterogeneity 
to the growth of the cell society.  The differentiation of
roles upon division reminds us of germ line segregation,
while multiple simultaneous deaths occurring repeatedly, 
reminds us of apoptosis. 
In this context, we propose that cellular interaction is 
relevant to the apoptosis observed in multicellular organisms.

The temporal oscillation of the population of different types of cells
in \S 5 may provide a basis for the understanding
of the experiment by Ko, Yomo, and Urabe \cite{Yomo},
where {\it E. coli} successively cultivated
in a well-stirred liquid culture differentiate into distinct types
whose populations  show a complex oscillation in the time course.

Interaction-dependent differentiation of cell types is also found
in tumor formation,
as shown in a series of experiments by Rubin \cite{Rubin}.
In the experiments, the cell-to-cell interaction is not global in
contrast with the experiment by Ko et al. Since
we have not included any spatially local
effects, our results  cannot be directly applied to their experiment.
In general, spatially local interactions
among cells are, of course, important as
development proceeds in a multicellular organism.
Indeed we have made some simulations of
a `short-ranged-coupling' version of our model.  So far the scenario
presented in the paper is still valid there.  First,
the clustering of the phase of oscillations supports the
time sharing for resources, and then
at a later stage, cells located closely to each other
start to be differentiated following the scenario in the present paper.  
At the next stage, due to the
local interaction, differentiated
cells start to become organized spatially, leading to the pattern formation.  
Simultaneous
death of cells appears again but in a spatially  localized manner.

To model the developmental process further, introduction of
cellular motions in real space will also be required.
Cells with internal dynamics move in real space according to their
interaction with surrounding elements.  Such a model
leads to a novel class of dynamical systems, called a coupled map gas and
is discussed in \cite{Shibata}.

As dynamical systems, the present study casts two novel
classes of models.  One is a dual coupled map, and the other is open chaos.

{\sl Dual coupled map}

In the model of \S 5, the dynamics can be regarded as a
dual coupled map, in the sense that each variable $x_n^m(i)$ on a
two-dimensional lattice $(m,i)$ is coupled over
two spaces, i.e., over all cells $i$ and  over chemical
species $m$.  Both the couplings are global, although the
former is all-to-all, and the latter is sparsely connected.
Since each globally coupled map shows clustering, the dual coupled map
leads to a competition of clustering with respect to 
each space (cell and chemical).  We have made several
simulations for 2-dimensional dual coupled maps with all-to-all
global interactions for each dimension\cite{KK-up}.  With the increase
of nonlinearity, first the selection of clustering in one
direction ( while retaining the synchronization against the other space) 
appears.
Then elements lose synchronization against both directions,
where the clustering involves both the dimensions, and is coded by 
a combination of the two spaces.
%With the further increase elements are desynchronized against both directions.
In the present context of a cell society, this combined clustering leads
to the differentiation of cells with different chemical characters
(roles).

%On digitalization

{\sl Open Chaos}

Our system, as a whole, has some kind of orbital instability
analogous with chaos.  
Within the internal dynamics, an orbit is expanded 
with the mechanism of chaos, and then is divided.
Although this expansion-division process looks like a Baker's transformation,
there is a clear difference.  After cell division, the dimension of
the phase space (i.e., the number of degrees of freedom) is increased, while in
a Baker's transformation the orbit stays in the
original phase space.  
We have coined the term ``open chaos" \cite{Rel} to address
such a chaotic instability which is inseparable from the
change of the phase space itself.

Due to the change of the number of degrees of freedom, most 
quantifiers for dynamical systems are not applicable, since
they require stationarity of the phase space.
In the present paper we have adopted the split exponent
to characterize the amplification rate of differences in the
internal states of two cells.
Since the exponent is defined locally at each time, it is an effective measure
in open ended dynamical systems.   
 
In a model allowing for the growth of the number of elements, 
we have found that the effective time sharing of system resources
is formed due to the presence of partially clustered states.  A balance between 
synchronization and desynchronization is necessary here, for the
effective use of resources, which enables the growth of the number
of cells (agents).  The sharing of resources by the clustering here
may be applied to economics, where a breakdown of
the time sharing system by synchronization may correspond to 
an economic crash.

{\sl acknowledgements}

I am grateful to Tetsuya Yomo and  Takashi Ikegami
for stimulating discussions, and to Brant Hinrichs for critical
reading of the manuscript.
The work is partially supported by
Grant-in-Aids for Scientific
Research from the Ministry of Education, Science, and Culture
of Japan.

\pagebreak

Figure Caption

Fig. 1

The overlaid time series of the fractional part of
$x_n(i)$.  When $x_n(i)$ exceeds 1, the line continues from
$x_n(i)-Int(x_n(i))$, with $Int(z)$ as the integer part of $z$.
The number of lines changes with division and death.
$T_g=10$, $T_d=0$, $s=0.1$.
(a) $K=2.0$ (b)-(d) $K=3.3$ (e) $K=4.0$.

\vspace{.1in}
Fig.2

Temporal evolution of the number of cells $N$.
 $T_g=10$, $T_d=0$, $s=0.1$.
for (a) $K=1.5$, 2.0, 2.5 and 
(b) $K=3.0$, 3.5, 4.0  (c) $K=3.3$.

\vspace{.1in}

Fig.3

The average (a) and maximal number (b) of cells
over the time steps 5000  to 105000.
Simulations are carried out with $s=0.1$, $T_d=0$,
and  starting from one cell. 
$T_g=100$ (solid line with $\Box$), 10 (dotted line with $\Box$), and 1 
(broken line with $\circ$).

\vspace{.1in}

Fig.4

Fraction of the average number of clusters to the total number of cells, i.e.,
$<k>/<N>$, obtained from the simulations for Fig.3.
$T_d=0$, and $s=0.1$.
$T_g=100$ (solid line with $\Box$), 10 (dotted line with $\Box$), and 1 
(broken line with $\circ$).

\vspace{.1in}

Fig. 5 

Split Exponent $\lambda_{spl}$ plotted with the change of $K$,
corresponding to Fig.3, and Fig.4.
$T_g=100$ (solid line), 10(dotted line), and 1(broken line).

\vspace{.1in}

Fig.6 

Histogram of the split exponent over a cell's lifecycle. $K=3.3$,
$s=0.1$, $T_g=10$, and $T_d=0$
(a) for  cells that divide successfully after their lifecycle
(b) for cells that die after their lifecycle.

Fig.7 

Split exponent $\lambda_{spl}(i,T_{birth},T_{division}$ over
the lifecycle plotted versus the lifetime of cells. $K=3.3$,
$s=0.1$, $T_g=10$, and $T_d=0$
(a)  for cells that divide successfully after their lifecycle
(b) for cells that die after their lifecycle.

Fig.8 

Temporal evolution of the number of cells $N$, plotted per $10^3$ steps.
$K=3.3$, $s=0.1$, $T_g=10$, and $T_d=0$.

Fig.9

Split exponent  $\lambda_{spl}(i),T_{birth},T_{division}$ over
the lifecycle for divided (a) and  dead (b) cells,
plotted in the time course  per $10^3$ steps.
The data are obtained from the simulations for Fig. 8.

Fig.10 

Cluster number  of  (a) divided and  (b) dead  cells, plotted
in the time course per $10^3$ steps.
The data are obtained from the simulations for Fig. 8.

Fig. 11

Cell lineage diagram corresponding to the simulation in Fig.8.
The vertical axis shows the time,
while the horizontal axis shows a cell index.  (For the
practical purpose of keeping track of the branching tree, we define
the index for the lineage
as follows: when a daughter cell $j$ is born from a cell $i$'s
$k$-th division, the value $s_j =s_i + 2^{-k}$ is attached to cell $j$
from the mother cell's $s_i$.
The index for cell $j$ is defined in the increasing order of $s_j$;
i.e., the index is sorted so that $s_j$ increases in the order of $j$.
Note that the idex for the lineage diagram
is different from the index adopted in all other figures, where
the cell index is given just as the order of birth).

In the diagram, a horizontal line shows the division of the cell 
of a smaller index producing the cell of a larger index,
while a vertical line is drawn as long as the cell exists 
(until it dies out).

Fig.12

Temporal evolution of the number of cells $N$
for the coupled circle map of model (4).
For Fig.12--15, we use the parameters
$K=1.1$, $c=0.59$, $s=0.8$, $T_g=1000$, and $T_d=-10$.
Simulations are carried out over $5\times10^7$ steps, 
starting from a single cell initial condition.
The number of cells is plotted per $10^4$ steps.
In the simulations, we have adopted 
$a^{\ell,m}=c$ for the pairs (1,5),(1,8),(2,5),(2,7),(3,1),(3,6),
(4,1),(4,2),(5,7),(5,8),(6,2),(6,4),(7,3),(7,4), (8,3),(8,6)
with $a^{m,\ell}=-a^{\ell,m}$, while other couplings are left to be zero.
The same behavior, though,  is observed for most other couplings 
satisfying with $L=2$.

Fig.13

Rotation $R^m_n(i)$ plotted against the cell index $i$.
The value with a mark gives $R^{(m)}_n(i)$ for
$m=1,2,\cdots 8$, for the cell with the index $i$.
Each line connecting between $R^{(m)}_n(i)$ is plotted 
only for the sake of visualization (i.e., to see the distinction by
chemical $m$ clearly).  Here cells that do not have
marks at the corresponding $i$ are already dead at the time.
(When cell $i$ dies, no cell with the index $i$ exists any more.
The cell index is attributed in the order of birth, without
compressing the indices for dead cells.)
For example,
cells with the indices 179,180 $\cdots,198$ have died at Fig.13(b.
Two lines with $R^m_n(i)>0$ correspond to the chemical species
$m=4$ and 6.

(a) time step $n=10^6$, when the cell number $N=33$, with the dead cell
of the index 32.
(b)  time step $n=8\times 10^6$
(c)  time step $n=20\times 10^6$
(d) time step $n=30\times 10^6$.

Fig.14 

Return map of the average rotation $R^1_{T_d}(i)$  over the lifecycle.
A daughter cell's average rotation over the lifecycle
is plotted versus its mother's cell's average before the division to
the daughter. 

Fig. 15
Rotation $R^1_{T_d}(i)$ over a lifecycle.
When a cell is divided its average rotation over the lifecycle
is plotted at the corresponding time step of the division. 
\end{document}